\documentclass{aa}
\usepackage{graphicx} \usepackage{natbib}
\usepackage[varg]{txfonts}
\newcommand{\HI}{{\ion{H}{i}}} 
\newcommand{\mJybeam}{mJy beam$^{-1}$} 
\newcommand{\WHz}{W Hz$^{-1}$} 
\newcommand{\msun}{{$M_\odot$}}
 
\newcommand{\kms}{km$\,$s$^{-1}$} 
\newcommand{\ltsima} {$\; \buildrel < \over \sim \;$} 
\newcommand{\gtsima} {$\; \buildrel > \over \sim \;$}
\newcommand{\lta} {\lower.5ex\hbox{\ltsima}} 
\newcommand{\gta} {\lower.5ex\hbox{\gtsima}}

 \usepackage{natbib} \usepackage{color,soul}
 \bibpunct{(}{)}{;}{a}{}{,} % to follow the A&A style
\newcommand{\atlas}{A{\small TLAS}$^{\rm 3D}$}

\begin{document}

\title{A rare example of low surface-brightness radio lobes in a gas-rich early-type galaxy: the
story of NGC~3998}

\titlerunning{Radio continuum and gas reservoir in NGC~3998} \authorrunning{Frank et al.}
\author{Bradley S. Frank\inst{1}$^,$\inst{2} \and Raffaella Morganti\inst{2}$^,$\inst{3} \and Tom
Oosterloo\inst{2}$^,$\inst{3} \and Kristina Nyland\inst{2}$^,$\inst{4} \and Paolo Serra\inst{5}}

\institute{Department of Astronomy, University of Cape Town, Private Bag X3, Rondebosch 7701, South Africa
    \and Netherlands Institute for Radio Astronomy (ASTRON), Postbus 2, 7990 AA Dwingeloo, 
            The Netherlands 
    \and Kapteyn Astronomical Institute, University of Groningen, Postbus 800, 9700 AV Groningen, 
            The Netherlands
    \and National Radio Astronomy Observatory (NRAO), 520 Edgemont Rd., Charlottesville, VA 22903, USA
    \and CSIRO Astronomy and Space Science, Australia Telescope National Facility, PO Box 76, Epping,
            NSW 1710, Australia}
        
\offprints{bradley@ast.uct.ac.za}

\date{Received ...; accepted ...}

\date{\today}

\abstract{We study the nearby lenticular galaxy NGC~3998. This galaxy is known to host a low-power radio AGN
with a kpc-size one-sided jet and a large, nearly polar \HI\ disc. It is therefore a good system to
study to understand the relation between the availability of cold-gas and the triggering of AGNs in
galaxies.  Our new WSRT data reveal two faint, S-shaped radio lobes extending out to $\sim$10~kpc
from the galaxy centre. Remarkably, we find that the inner \HI\ disc warps back towards the stellar
mid-plane in a way that mirrors the warping of the radio lobes. We suggest that the polar \HI\ disc
was accreted through a minor merger, and that the torques causing it to warp in the inner regions
are also responsible for feeding the AGN. The ``S'' shape of the radio lobes would then be due to
the radio jets adapting to the changing angular momentum of the accreted gas. The extended radio
jets are likely poorly collimated, which would explain their quick fading and, therefore, their
rarity in galaxies similar to NGC 3998. The fuelling of the central super-massive black hole is
likely occurring via ``discrete events'', suggested by the observed variability of the radio core and
the extremely high core dominance, which we attribute to the formation and ejection of a new jet
resulting from a recent fuelling event.}


\maketitle
%--------------------------------------------------------------------
\section{Introduction} 
Radio Active Galactic Nuclei (AGN) are an important sub-population of the general family of AGN
which show a variety of different properties and morphologies.  The low radio power group
\citep[$P_{\rm 1.4~GHz} < 10^{23}$ \WHz;][]{Sadler1989,Wrobel1991a,Nyland2013}  form the bulk of
the radio emitting AGN population \citep{Sadler2007,Best2005}. Despite their low power, they can
impact the ISM of the host galaxy as shown in a number of cases, e.g. NGC~1266
\citep{Alatalo2011,Alatalo2015,Nyland2013,Nyland2016}, NGC~1433 \citep{Combes2013} and  IC~5063
\citep{Morganti2015}. It is therefore important that we study this population in more detail. 
Compared to powerful radio galaxies, low power radio sources tend to be more core dominated and have
smaller spatial extents \citep{Nagar2005,Baldi2010,Baldi2015,Nyland2016}. The low radio power group
($P_{1.4~GHz} < 10^{23}\,\mathrm{W\,Hz^{-1}}$) \citep{Sadler1989,Wrobel1991a,Nyland2013} form the
bulk of the radio emitting  AGN population \citep{Best2005,Sadler2007}. Although still poorly
understood, the brightness, morphology and scale of the radio emission may be related to a lower, or
inefficient, supply of fresh gas or to weak/poorly collimated jets likely due to lower flow
velocities and, therefore, are more subject to instabilities and entrainment
\citep{Nagar2005,Laing2011,Baldi2015}.

Considering the likely and widespread availability of gas, an important question is what the
mechanisms are that can bring the gas to the inner regions and trigger the super-massive black hole
(SMBH). Mergers are often suggested to perform this task. However, large time delays are often seen
between the merger/accretion event and the onset of the radio source
\citep[e.g.][]{Emonts2006,Struve2010a,Shulevski2012,Maccagni2014}. This seems to rule out a direct
link between the two processes and indirect links should instead be considered \citep[see,
e.g.,][]{Wild2010}. An alternative is the fuelling  by  gas clouds condensed from the hot halo, or
from an available reservoir of cold gas near the centre, as has been suggested by a number of
authors \citep[e.g.][]{Allen2006,Hardcastle2007,Gaspari2015}. Evidence  of such clouds  in the
process of fuelling the central SMBH has been found in the case of the young radio source
PKS~1718--63 \citep{Maccagni2014,Maccagni2016}.  Finally, the combination with secular processes is
also likely to play a role.

Here we present the results of deep radio continuum and \HI\  observations of the nearby, low radio
power and  gas-rich early-type galaxy, NGC~3998. The observations reveal a particularly intriguing
and uncommon  structure of the radio continuum emission. The presence of the \HI-rich structure
allows to explore the connection between the radio lobes  and the large reservoir of gas in the
system.
%------------------------------------------------------------------------
\section{Overview of NGC 3998} 

NGC~3998 is a nearby\footnote{In this paper we assume a distance of of 13.7 Mpc for NGC~3998
\citep{Cappellari2011a}. At this distance, 1$^{\prime\prime}$ corresponds to  65.9 pc.} S0 galaxy
located at the outskirts of the Ursa Major group and recently studied in detail as part of the
multi-wavelength ATLAS$^{\mathrm{3D}}$ survey \citep{Cappellari2011a}.  This galaxy hosts a
low-power, $\sim 10^{22}$ \WHz, radio  AGN \citep{Hummel1980,Wrobel1984,Wrobel1991}  with a  flat
spectrum radio core \citep{Hummel1980,Hummel1984,Kharb2012}. Very long baseline interferometry
(VLBI) observations have revealed a jet-like structure on the northern side of the nucleus of NGC
3998 \citep{Filho2002,Helmboldt2007}. The radio core is known to be variable (see also Sec. 3.1) and
NGC~3998 also shows variability  in the ultraviolet \citep{Maoz2005} and X-ray
\citep{Younes2012,Hernandez2013} regimes.  At X-ray energies, NGC\,3998 contains a nuclear source
with $L_{\mathrm{X}}$ (2-10\,keV) $\sim$ $2.5\times10^{41}\,\mathrm{erg\,s^{-1}}$
\citep{Terashima2000,Ptak2004}.

The optical spectrum of NGC~3998  shows broad H$\alpha$ emission lines
\citep{Heckman1980,Devereux2011} although it is formally classified as a LINER 1.9 \citep{Ho1997}.
Imaging studies of the ionised gas in this galaxy have revealed a small H$\alpha$ disk of about 100
pc \citep{Pogge2000} surrounded by a larger ($\sim 5$ kpc), warped structure
\citep{Ford1986,Sanchez2012}. The nuclear region of NGC~3998 is also known to contain dust with an
estimated mass of $M_{\mathrm{dust}} \sim 5 \times 10^5$ \msun\ \citep{Martini2013}.
In terms of its stellar population, NGC~3998 is a red galaxy $(u-r)=4.46$ (SDSS DR6,
\citealt{Adelman2008}) dominated by evolved stars \citep{McDermid2015}.  The
single-stellar-population age inside one effective radius ($\sim$ 20$^{\prime \prime}$) reported in
\citet{McDermid2015} is 11.48 $\pm$ 2.00 Gyr.  However, young stars (with ages of about 1--10 Myrs)
may dominate the nuclear ($< 1^{\prime \prime}$) region of NGC~3998
\citep{Gonzalez2004,Gonzalez2009,Mason2015}.

NGC~3998 contains a large, spatially extended reservoir of H\,{\sc i} arranged in a disk-like
structure \citep{Knapp1985}. More recently, \citet{Serra2012} showed that the \HI\ disk is many
times larger than the optical component and has a mass of $2.8\times10^8$ \msun.  Studies of the
\HI\ content of galaxies in Ursa Major \citep[][and references therein]{Pak2014} have noted the
presence of tidal gas around NGC~3998, which is also seen in the \citet{Serra2012} results. This is
likely connected to the fact that NGC~3998 is located in a region of fairly high galaxy density
\citep{Cappellari2011b}.   CO observations by \citet{Baldi2015} revealed that NGC~3998  hosts
molecular gas with an H$_2$ mass of $\sim$ 1.7 $\times$ 10$^7$ \msun. 
%------------------------------------------------------------------------
\section{Observations and Data Reduction}

\begin{table} 
    \caption{\label{tab:obs}Summary of WSRT 21-cm narrow-band observations.} 
    \centering
    \begin{tabular}{l l} 
        \hline 
        \hline 
        $N_{dishes}$ & 11-12\\
        Total observing time & 108 h\\
        Observing period & June 2009 -- May 2011 \\ 
        Bandwidth/channels & 20~MHz/1024 \\
        Velocity resolution & 4.2 $\mathrm{km\,s^{-1}}$\\
        Continuum resolution & $16.5''\times13.2''$, P.A.$=2.4^{\circ}$\\ 
        3-$\sigma$ noise continuum image & 0.15 \mJybeam \\ 
        3-$\sigma$ noise per channel ($15''$ cube) & 0.13 \mJybeam \\
        \hline 
    \end{tabular} 
\end{table}

\begin{table} 
    \caption{\label{tab:obs-wb}Summary of WSRT broad-band observations.} 
    \centering
    \begin{tabular}{l l} 
        \hline 
        \hline
        $N_{dishes}$    & 5\\
        Observing time per track & 12 h\\
        Subbands/Bandwidth/channels & 8$\times$20~MHz/128 \\
        \hline
        \em{21-cm Observations}\\
        Date Observed & 2 June 2015\\
        Band Centres [MHz] & 1450,1432,1410,1392,\\
                &       1370,1350,1330,1311\\
        3-$\sigma$ noise continuum image & 0.43 \mJybeam \\ 
        Continuum resolution & $16.5''\times12.1''$, P.A.$=85.8^{\circ}$\\ 
        \hline
        \em{6-cm Observations}\\
        Date Observed & 20 June 2015 \\
        Band Centres [MHz] & 4858,4876,4894,4912,\\
                            & 4930,4948,4966,4984\\
        3-$\sigma$ noise continuum image  & 0.41 \mJybeam \\ 
        Continuum resolution & $7.00''\times4.966''$, P.A.$=52.2^{\circ}$\\ 
        \hline 
    \end{tabular} 
\end{table}

We conducted two campaigns to observe NGC~3998 with the Westerbork Synthesis Radio Telescope (WSRT).
The observations  performed with a relatively narrow (spectral line) band were focused on the study
of the \HI, but have also provided  high-quality radio continuum images.  In 2015, we further
conducted broad-band observations at wavelengths of 21-cm and 6-cm to better estimate the  flux
variation of the radio continuum.  Below we provide an overview of the calibration and imaging
procedure for each set of observations. For the data reduction of both observations, we have used
the \texttt{MIRIAD} \citep{Sault1995} package to perform the data calibration and imaging.

\subsection{Narrow-Band Observations}

The spectral line observations were conducted as a deep follow-up to those presented in
\citet{Serra2012}  which were part of the $\mathrm{ATLAS^{3D}}$ survey. We obtained  nine 12-hour
tracks during the period from 2009 to 2011 and a summary of the observational parameters  is given
in Table\ref{tab:obs}.  The number of dishes that were available during this period ranged
between 11 and 12. This was due to the permanent absence of one dish to the DIGESTIF project
\citep{5171752}, and the mechanical maintenance of several other dishes in preparation for APERTIF
\citep{2009wska.confE..70O}.
Doppler tracking on a heliocentric velocity of 1000 \kms\ was used during the observations. The
bandwidth of each observation was 20 MHz, distributed over 1024 channels giving a channel width of
4.2 \kms. In making the datacubes, the data were binned to a velocity resolution of 8.4 \kms.  

The initial part of the calibration followed the standard s.pdf such as flux and bandpass
calibration.  We then inspected the target spectrum and identified which channels were free of line
emission. We then created line and continuum visibility datasets by doing a continuum subtraction,
using a fourth-order polynomial to model the bandpass over the line-free channels.  We performed
phase-only self-calibration on the continuum dataset and the calibration tables derived in this way
were also copied to the line dataset.  This process produced continuum images which were dynamic
range limited due to direction-dependent errors around a bright continuum source at the edge of the
field of view. Such direction dependent errors are due to a combination of small pointing errors and
the effects of standing waves in the optics of the WSRT dishes.  We solved this issue by doing a
direction dependent calibration known as
``peeling''\footnote[1]{\url{http://www.astron.nl/~oosterlo/peeling.pdf}}. 

The spectral line images were made using the Robust weighting scheme \citep{Briggs95} with a
robustness of 0.4 and using three different tapering schemes, corresponding to spatial resolutions
of $15''$, $30''$ and $60''$. We then cleaned each cube by the iterative use of threshold masking
(on the cleaned and residual images). The high-resolution cubes we use for examining the kinematics
of the \HI, while the low-resolution cubes we use for studying the properties of the extended, low
surface brightness gas. 

We then imaged the combined continuum visibility data using uniform weighting and cleaned the
continuum image using a similar method of iterative masking and cleaning that we used to image the
line data. 

\subsection{Broad-Band Observations}

The second observing campaign comprised follow-up radio continuum observations at 21-cm and 6-cm in
June 2015. The relevant information for the broad-band observations are summarized in Table
\ref{tab:obs-wb}.  The purpose of these observations was to do broad-band measurements of the
variable radio continuum emission of the AGN.  Two 12-hour tracks were conducted at  6- and 21-cm,
respectively.  Only  five WSRT dishes  were available for these observations due to the ongoing
upgrade of the WSRT to the phased-array-feed based APERTIF system.  For each observation we used the
standard 160 MHz bandwidth setups available with the WSRT.

Each of the 20~MHz sub-bands were covered by 128 channels in dual polarization. Standard flux and
bandpass corrections were derived from short observations of 3C48   and were copied  to the target
visibilities. We then performed  iterative self-calibration on each sub-band in the same way as
described in the previous section. The final images were produced using uniform weighting.

\begin{figure}
    \centering 
    {\includegraphics[width=\hsize]{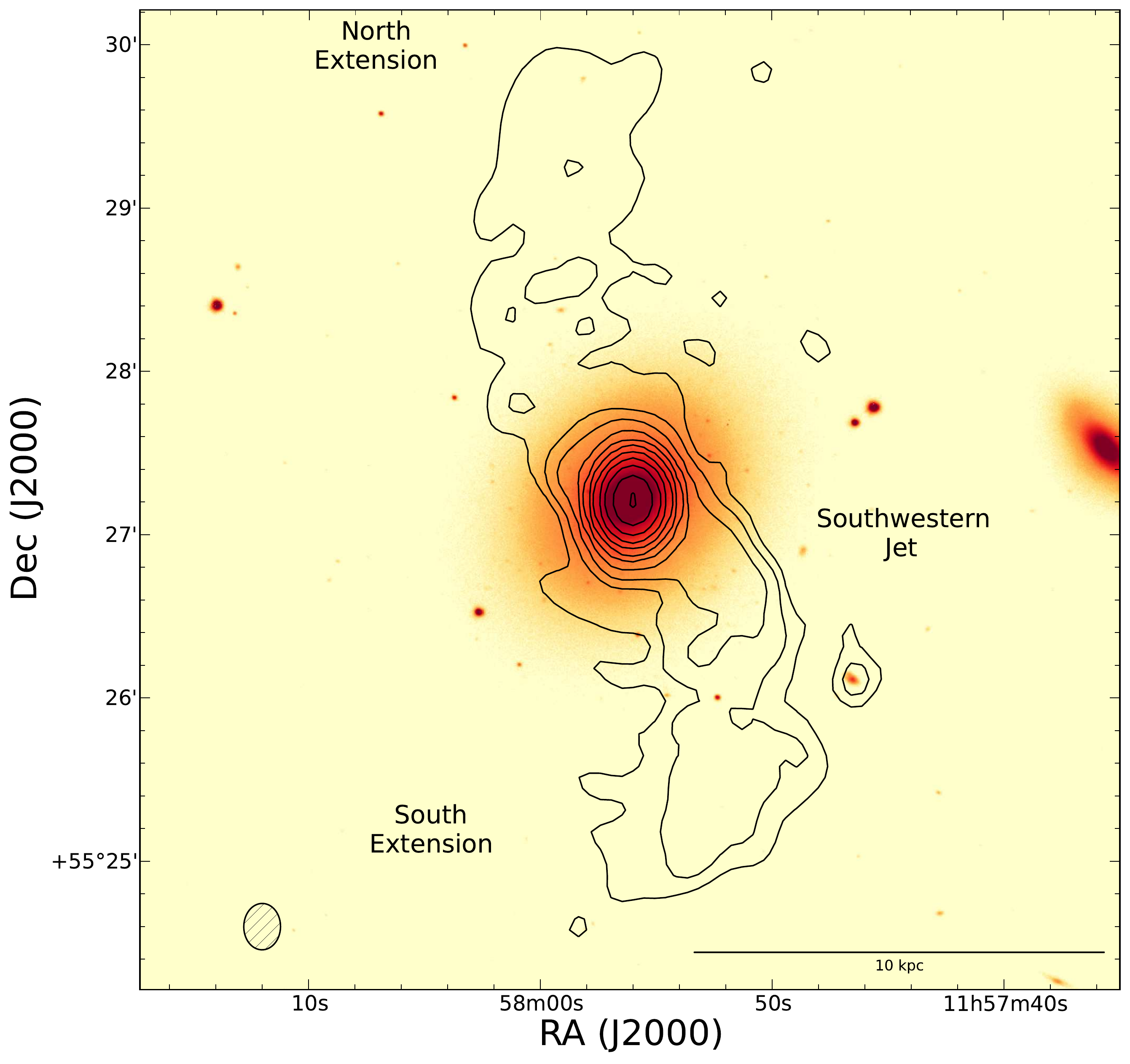}}
    \caption{\label{fig:cont} Radio continuum contours from the narrow-band 21-cm of NGC~3998
        overlaid on an {\em r}--band optical image from the Sloan Digital Sky Survey
        \citep{2015ApJS..219...12A}. The contours from the radio continuum were produced from
        the 21-cm narrow-band data using uniform weighting --- which corresponds to a robustness
        parameter in \texttt{MIRIAD} of --2. The beam-size is indicated in the lower-left corner
        of the image. The contours are given by $S_{\rm c}=S_0\times2^n\,\mathrm{mJy\ beam}^{-1}$, for
        $n=0,1,2,...,10$, where $S_0=0.150\,\mathrm{mJy\ beam^{-1}}$ is the 3-$\sigma$ sensitivity as
        indicated in Table \ref{tab:obs}.} 
\end{figure}

\begin{figure} 
    \centering 
    {\includegraphics[width=\hsize]{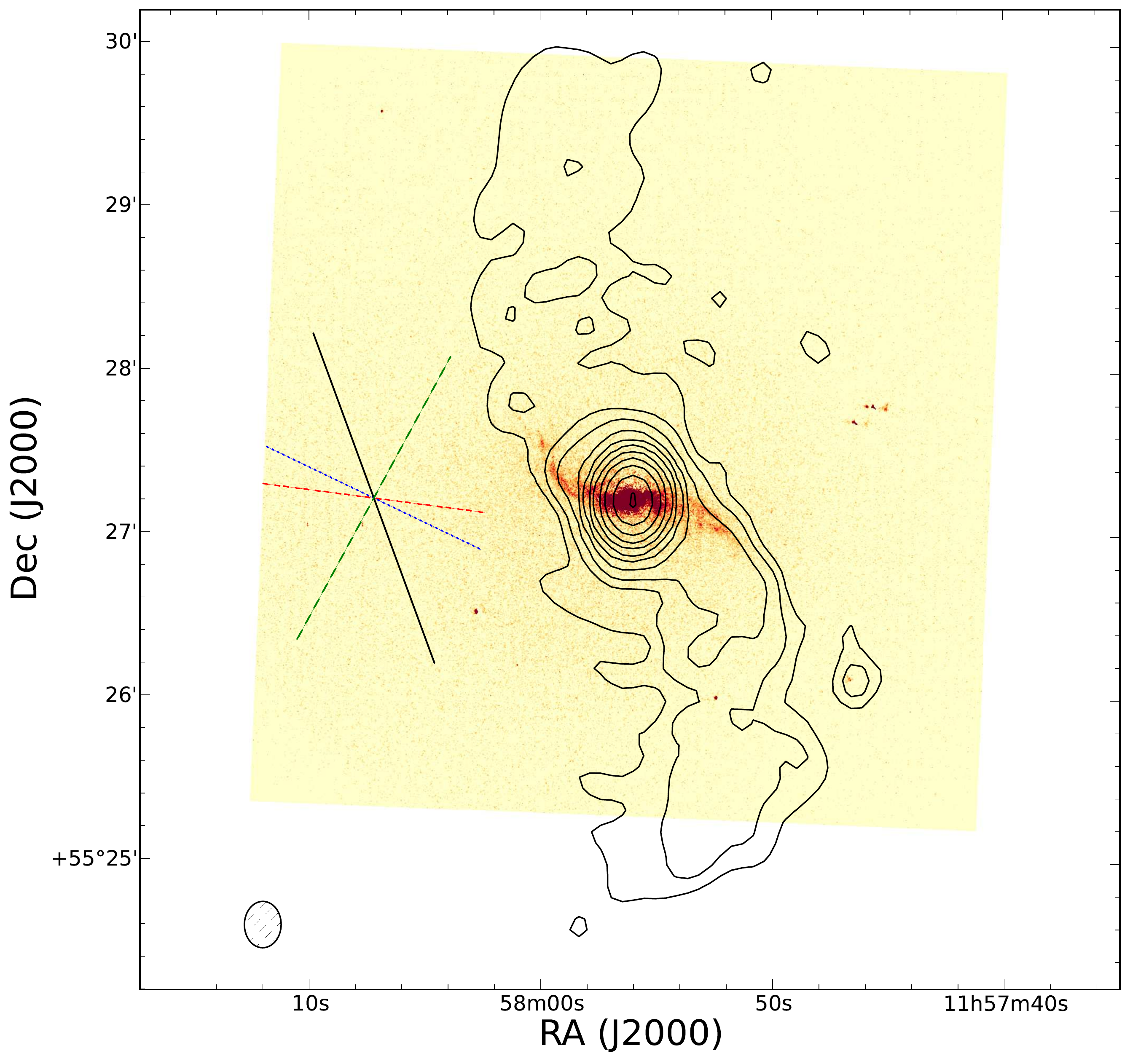}} \caption{\label{fig:ngc3998-ha} Radio
        continuum contours from the narrow-band 21-cm of NGC~3998 overlaid on the
        continuum-subtracted $\mathrm{H\alpha}$ image from \citet{Sanchez2012}. The contours are
        described in Fig.  \ref{fig:cont}.  On the left we show a depiction of the position angles of
        the various galaxy components: green-dashed: stars (PA$\sim$$135^\circ$, from
    \citealt{Krajnovic2011}), black-solid: inner $\sim$5 kpc of radio continuum
(PA$\sim$$33^\circ$), blue-dotted: \HI\ disk (PA$\sim$$65^\circ$), red-dot-dash: H$\alpha$ disk
(PA$\sim$$84^\circ$).  } 
\end{figure}

%--------------------------------------------------------------------
\section{Results}
%--------------------------------------------------------------------
\subsection{The Radio Continuum}

In Fig.\ref{fig:cont} we present the radio continuum image of NGC~3998 as derived from the
narrow-band observations. As indicated in Table \ref{tab:obs-wb} and the previous section, less than
half of the WSRT array was available for the broad-band observations conducted in 2015. As a result,
the 6-cm observations are not sensitive to any diffuse, faint structure, and we only detect the
unresolved core. Similarly, the broad-band observations at 21-cm are less sensitive than the
narrow-band observations.
Assuming the flux density varies as a function of frequency $S\propto\nu^{-\alpha}$, we calculate a
spectral index $\alpha$ from the 21- and 6-cm broad-band observations for the core of 0.19, which is
flat and characteristic of synchrotron self-absorption commonly associated with a radio AGN. The
rest of the analysis and discussion in this paper focuses on the 21-cm narrow-band emission, unless
otherwise explicitly stated.

The narrow-band continuum emission appears particularly interesting and is much more
extended than was previously known.  The new observations reveal two  poorly collimated, low-surface
brightness structures extending  N and S of the core. The projected linear extent of each of these
lobes  is $\sim$10 kpc.  A hint for the presence of the   southern (and brighter)  lobe is visible
on the NRAO VLA Sky Survey \citep[NVSS,][]{1998AJ....115.1693C} images and was also earlier  reported
by \citet{Wrobel1984} and  \citet{Knapp1985}.

The ``S-shaped'' large-scale morphology of these lobes is immediately apparent from Fig.\
\ref{fig:cont}. Interestingly, the shape of the radio continuum appears to mirror the warped
structure of the  ionised gas  disk \citep{Sanchez2012} as illustrated in Fig.\ \ref{fig:ngc3998-ha}
We discuss this in more detail below.  In the central region, we observe a bright  core while the
brighter part of the SW jet structure extends to $\sim$1.5 kpc from the nucleus.  Interestingly, the
VLBI (Very Long Baseline Array, VLBA, and European VLBI Network, EVN) observations reported by
\citet{Helmboldt2007} and  \citet{Filho2002} show an extension on the scale of $\sim$10 mas to the
north of the core. The connection between the mas and the arcsec  scale will need to be investigated
with new observations.

We measured integrated flux densities from the radio continuum map using the Viewer from the Common
Astronomy Software Applications (CASA)\footnote{http://casa.nrao.edu} by drawing a polygon around
the region of interest; each polygon large enough to encompass emission brighter than approximately
3-$\sigma$ as indicated in Table \ref{tab:obs}. For the core emission we fitted a Gaussian shape
to the unresolved emission. 

Assuming a 5\% uncertainty in the absolute flux scale at WSRT, we calculated the total error as the
quadratic sum of the theoretical uncertainty over the region of interest and the systematic
uncertainty associated with the measured flux. 

The integrated flux densities, errors and the associated luminosities are presented in Table
\ref{tab:cont}. The integrated flux density of the northern lobe is 7.1~mJy while the southern lobe
has a slightly higher flux for the extended part (11.6~mJy). For the bright, linear extension
directly to the southwest of the core we derive a flux density of approximately 3.3 mJy. 

In addition to the S-shape of the continuum emission,  two other interesting properties can be
noted: the relatively low surface brightness of the extended structures and the large core
dominance.  The newly discovered lobes have a surface brightness of about 5 mJy arcmin$^{-2}$. This
is low and at the limit of what is detected with currently available surveys and indeed only the
southern lobe of NGC~3998 is barely visible in the NVSS image. Structures with low surface
brightness radio continuum emission are often considered to be a signature of remnant-like emission,
i.e.\ emission from lobes that are in the process of fading away because they are not supplied
anymore with fresh radio plasma (e.g. \citealt{Saripalli2012,Brienza2016}).  Furthermore, the
detected radio structure is strongly core dominated. The ratio $S_{\rm core}/S_{\rm ext} \sim
8$ is much higher than what is typically found  for  FRIs \citep{Morganti1997,Laing2014} and is
almost an order of magnitude higher than   in even the   radio sources of even lower power
considered by \citet{Baldi2015}. We will return to these points later in the discussion.

Based on the narrow-band observations, we measure a radio continuum flux density of 176.7 mJy for
the core emission.   This value is much higher than  measured from the NVSS image  (106.8 mJy) and
by  \citet{Knapp1985}, $\sim$100 mJy.  In order to check whether our measurement could be due to an
error in the flux calibration, we compared our measurements of the flux densities of other sources
in the field with those in the NVSS image. We found the fluxes measured from our images to be within
$5\%$ of the values measured from the NVSS image. 

\begin{figure} 
    \centering 
    {\includegraphics[width=\hsize]{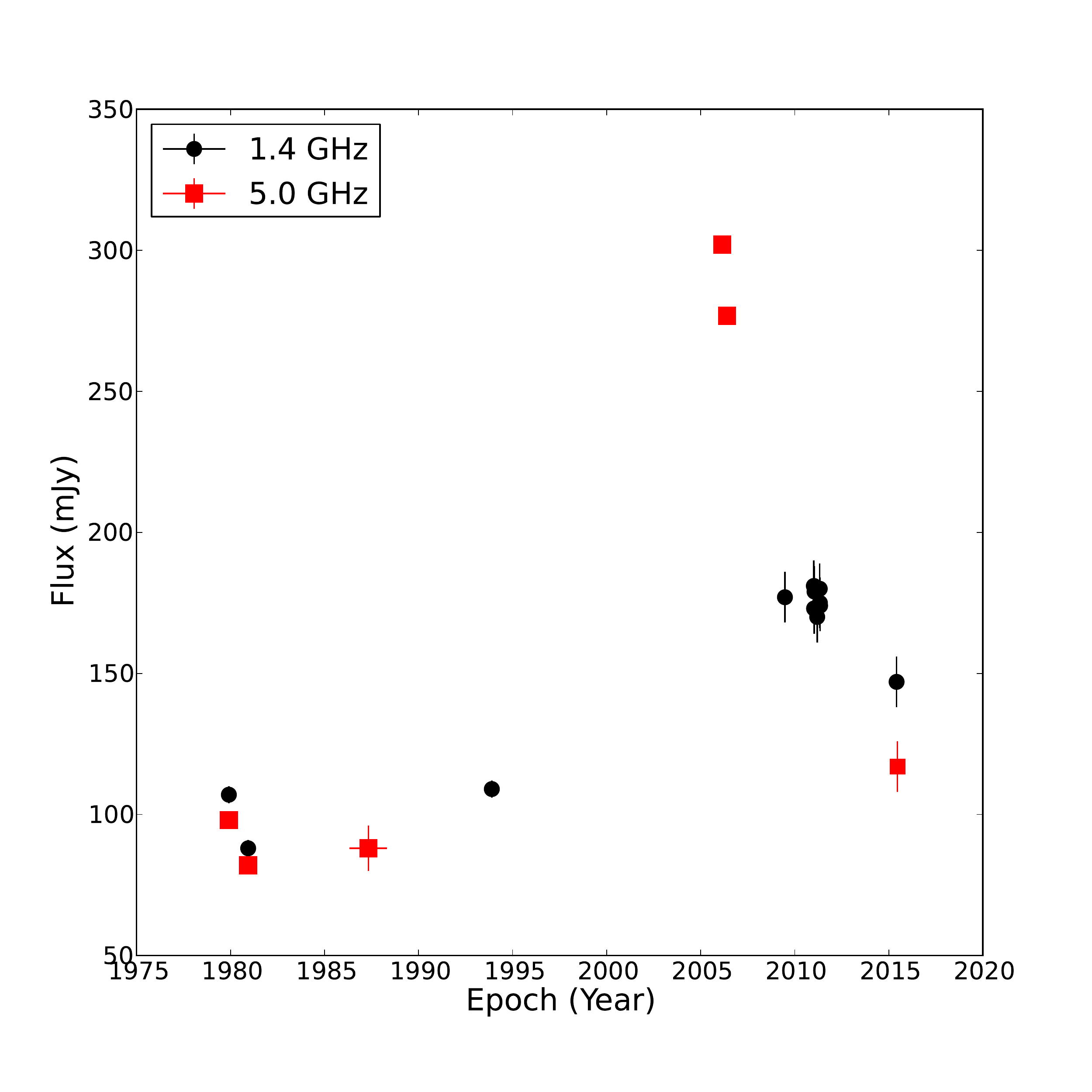}}
    \caption{\label{fig:fluxtime} Variability of the radio continuum emission from the core of
    NGC~3998. The red squares correspond to measurements at 6-cm, the black circles at 21-cm.
    The epochs, measurements and associated uncertainties (where available) are presented in Table
    \ref{tab:var}} 
\end{figure}

Although the radio core of NGC~3998 is known to be variable, the flux density of NGC~3998 did not
show any time variation for the duration of the narrow-band campaign, i.e.,  the flux density of the
core  remained constant from 2009 to 2011.   However, when considered over a longer time span, it is
clear that the radio core varies substantially in flux density. In Fig.\
\ref{fig:fluxtime}\footnote{These measurements were derived using different telescopes, with
different resolutions/UV-coverages. We expect that this will have a small effect on the observed
variability depicted in the figure.}  we compare our measurements with those from   the literature
as a function of time. We also list the measurements used to construct this plot in Table
\ref{tab:var}, with the associated uncertainties where available. The flux reported from the 2006
VLBA observations at 6-cm of \citet{Helmboldt2007} is 276.8 mJy, suggesting that our observations
may have been during  a phase of decreasing flux after a flare. This is consistent with  our
follow-up broad-band radio continuum observations of NGC~3998 conducted during 2015, which have
yielded a flux density for the core emission of 148 mJy at 21-cm and 117 mJy at 6-cm.
\citet{Kharb2012} reported their 2006 measurement of the 4.9 GHz flux density of 302 mJy which is
more than three times higher than the measurement presented in \citet{Wrobel1991} of 83 mJy.
\citet{Gregory1996} presented a 4.85 GHz flux density of 88 mJy, as measured with the Green Bank
telescope during the observational campaign in 1986.  We measure a 5 GHz flux density of 117 mJy.
Thus, also at this higher frequency, the core flux of NGC~3998 appears to be declining after a
recent burst. 

\begin{table} 
    \caption{\label{tab:var}Flux Variability of NGC~3998. We present the epoch, frequency, flux density
    and uncertainty (where available) and the associated reference for the measurement.} 
    \centering
    \begin{tabular}{l l r c} 
        \hline 
        \hline 
        Date & Frequency & Flux Density & Reference \\
             &  MHz & mJy &  \\
        \hline 
        30/11/1979\tablefootmark{a} & 1500 & $107\pm3$ & (1)\\
        30/11/1979\tablefootmark{a} & 4900 & $98\pm3$ & (1)\\
        09/12/1980                  & 1500 & $98\pm3$ & (1)\\
        09/12/1980                  & 4900 & $83\pm3$ & (1)\\
        02/05/1987\tablefootmark{a} & 4850 & $88\pm8$ & (2)\\
        23/11/1993                  & 1420 & $109\pm3$& (3)\\
        24/02/2006                  & 4860 & 302 & (4)\\
        27/05/2006                  & 4844 & 277& (5)\\
        25/06/2009 & 1420 & $177\pm9$ & This work.\\
        15/02/2011\tablefootmark{b} & 1420 & $175\pm9$ & This work.\\
        02/06/2015 & 1381 & $149\pm7$ & This work.\\
        20/06/2015 & 4901 & $118\pm6$ & This work.\\
        \hline 
    \end{tabular} 
    \tablebib{
        (1) \citet{Wrobel1984};
        (2) \citet{Gregory1996};
        (3) \citet{1998AJ....115.1693C};
        (4) \citet{Kharb2012};
        (5) \citet{Helmboldt2007}
    }
    \tablefoot{
        \tablefoottext{a}{In cases where there was an ambiguity in the date
        (e.g., survey data/several epochs), the median date was used.}
        \tablefoottext{b}{We show the median date and flux density deduced from our 2011
        observations.}
    }
\end{table}

% Table with Luminosity in erg/s
\begin{table} 
    \caption{\label{tab:cont} Flux density and radio luminosity for the continuum features
    denoted in Fig.\ \ref{fig:cont}.} 
    \centering 
    \begin{tabular}{l c c } 
        \hline 
        \hline 
        Continuum feature & $S_{\mathrm{1.4~GHz}}$ & $L_{\mathrm{1.4~GHz}}$   \\ & mJy & $10^{31}\,\mathrm{erg\,s^{-1}}$\\
        \hline 
        Total & $195.6\pm10.3$ & 61.5\\ 
        Core & \phantom{2}$176.7\pm8.9$ & 55.6 \\ 
        Northern extension & \phantom{222}$7.1\pm1.5$ & \phantom{2}2.2\\ 
        Southern extension & \phantom{22}$11.6\pm1.4$& \phantom{2}3.6\\ 
        Southwestern jet  &  \phantom{222}$3.3\pm0.3$& \phantom{2}1.1 \\ 
    \hline 
\end{tabular} 
\end{table}

%--------------------------------------------------------------------

\subsection{Atomic Gas in NGC~3998}

\begin{figure*}
    \centering 
    \includegraphics[width=\hsize]{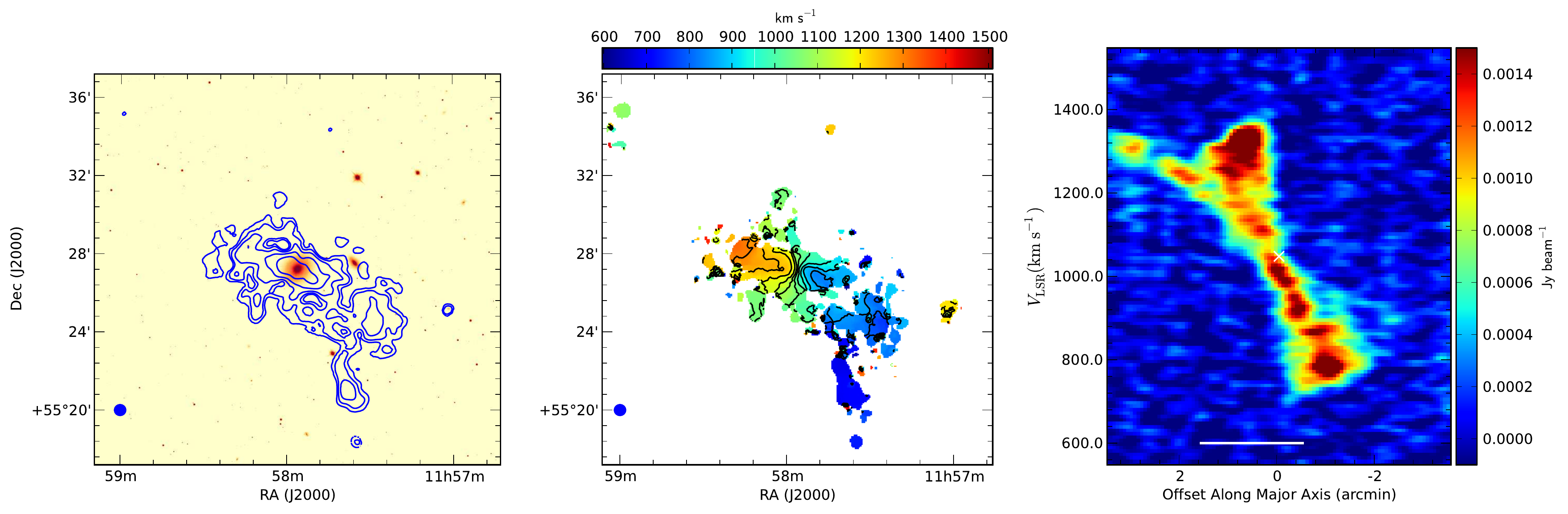}
    \caption{\label{fig:ngc3998-hi} Left: \HI\ column density contours, calculated from the $30''$
        cube, overlaid on the {\em r}-band SDSS image. The contours are given by  $S_\mathrm{H
        \textsc{i}} = 2^n  \times 10^{19} \mathrm{cm^{-2}}\ (n=0, 1, 2, ...)$ The beam size is
        indicated in the lower-left corner of the figure. Middle:
        Intensity-weighted mean velocity field calculated from the $30''$ cube, the iso-velocity
        contours correspond to $V_\mathrm{sys}\pm n\times50 \mathrm{\,km\,s^{-1}}$, where $n=1, 2, 3,
        ...$ and $V_\mathrm{sys}=1048\mathrm{\,km\,s^{-1}}$.  Right: Position-velocity diagram along the
        major axis of the $15''$ \HI\ cube.  The white scale bar indicates 10 kpc.}
\end{figure*}

\begin{figure*} 
    \centering 
    \includegraphics[width=14cm]{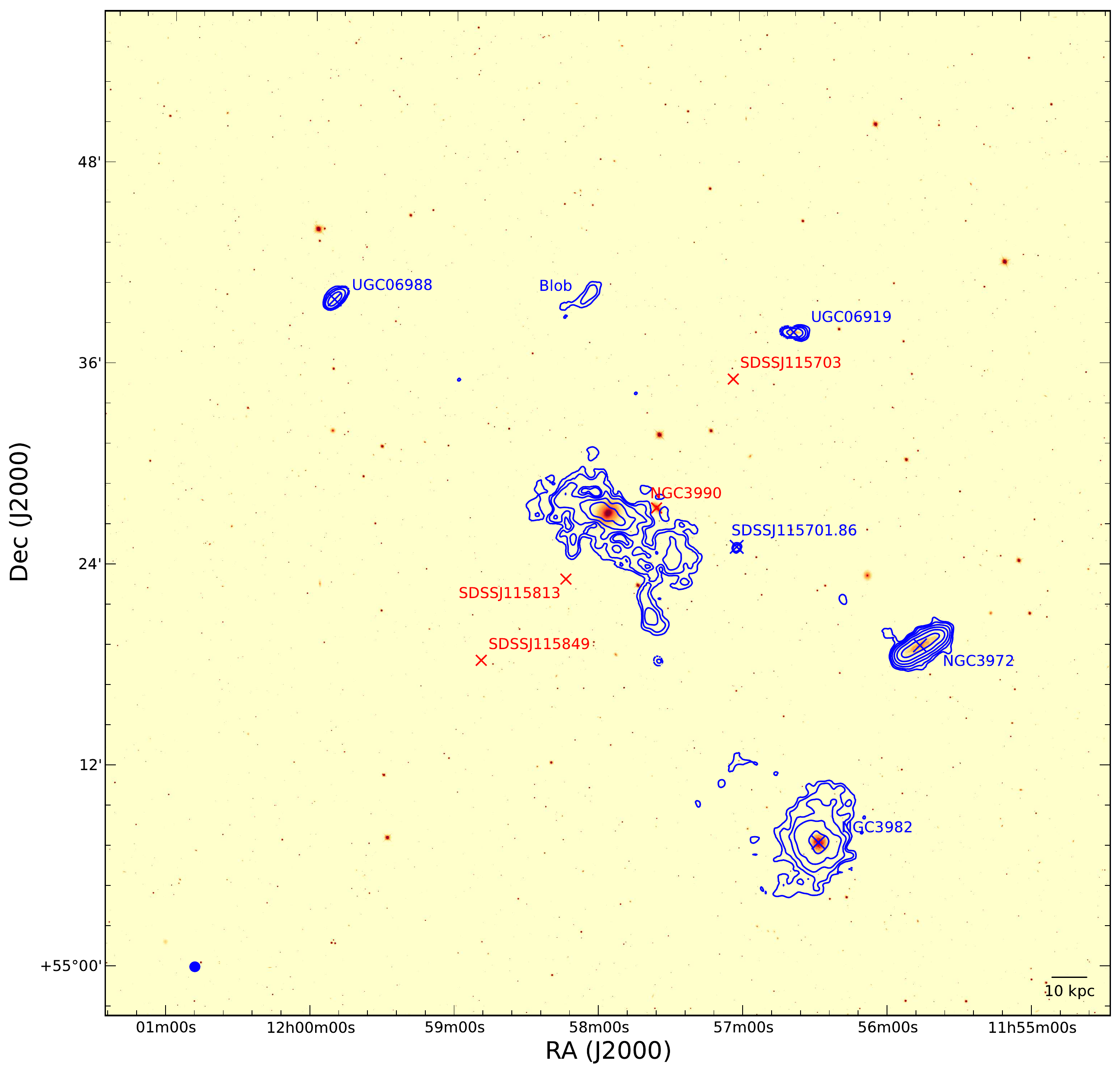}
    \caption{\label{fig:hi-full}H\,{\sc i} column density contours of NGC~3998 and its neighbours,
        overlaid on the {\em r}-band SDSS image.  The contours are given by $S_\mathrm{H \textsc{i}}
        = 2^n  \times 10^{19} \mathrm{cm^{-2}}\ (n=0, 1, 2, ...)$; the lowest contour corresponds to
        the detection limit of our observations. The Figure is annotated such that blue crosses and
    labels indicate galaxies with an H\,{\sc i} detection in our work, while a red cross and labels
indicates galaxies that do not have an associated detection. } 
\end{figure*}

In Fig.\ \ref{fig:ngc3998-hi} we present the \HI\ column density map and velocity field derived from
the  $30''$ image cube and the major axis position-velocity (PV) diagram from the  $15''$ data.  The
most striking feature of the \HI\ in NGC~3998 is that much of the \HI\  is arranged in a highly
inclined disk that is roughly polar with respect to the distribution of the stars while at larger
radii the \HI\ morphology is less regular. In Fig.\ \ref{fig:hi-full} we present an image of
the \HI\ emission in NGC~3998 and the neighbouring galaxies. The emission in NGC~3998 is lopsided
and low-column density gas extends to the S-W --- towards NGC~3982 and NGC~3972. Additionally,
there are streams and trails of low column density gas which indicates tidal interactions
between the three most massive galaxies in this group. The structure of the \HI\ could suggest
that galaxy interactions  have occurred and that (some of) the  \HI\ gas found in NGC~3998
originates  from tidal interactions with other members of the group, similar to the distribution
observed around M81 \citep{1994Natur.372..530Y} (also see \citealt{1984ARA&A..22..445H} and
\citealt{2001ASPC..240..657H} for more examples of interacting systems). 

NGC~3998 is part of a group of galaxies at the edge of the Ursa Major cluster \citep{Tully1996},
several of which are detected by our observations (Fig.\ \ref{fig:hi-full}). The 5-$\sigma$
detection limit of our observations is about $5\times10^5$ \msun for a velocity width of 8.4 \kms.
In Table \ref{tab:HI} we provide the integrated flux densities, the associated uncertainties and the
\HI\ masses for each of the detections, as measured using the $30''$ cube, which was corrected for
the primary beam.  When calculating the \HI\ for each object we used the distance to NGC~3998 of
13.7 Mpc, which fits in with our interpretation of these objects being involved in inter-group
interactions. This will be discussed in the Sections below.

From our data we derive a total \HI\ mass of the \HI\ disk in NGC~3998 of $4.3 \times 10^8$ \msun,
which is considerably higher than the value reported by \citet{Serra2012}. This is likely due to the  increased
detection of low column density \HI\ in our deeper observations.  The H\,{\sc i}  mass determined
for NGC~3982 is close to the value mentioned in \citet{Boselli2014}, and the H\,{\sc i} mass for
NGC~3972 is similar to the value mentioned in \citet{Sanders1998}.  SDSSJ115701 is the only 
other SDSS object for which we have a clear H\,{\sc i} detection. 

\begin{table}[b] 
    \caption{\label{tab:HI}Integrated flux densities and \HI\ masses for the galaxies detected in
    our observations.}
    \centering 
    \begin{tabular}{l r c c c c} 
        \hline 
        \hline 
        Galaxy 	        & Integrated Flux Density       & H\,{\sc i} Mass\\ 
                        & $\mathrm{Jy\,\,km\,s^{-1}}$   & $\log M_\mathrm{H\textsc{i}} / {M_\odot}$\\
        \hline 
        NGC~3998        & $9.83\pm0.13$                 & 8.63 \\
        NGC 3972        & $15.96\pm0.08$                & 8.81\\ 
        UGC 6988        & $1.23\pm0.19$                 & 7.74\\ 
        UGC 6919        & $0.76\pm0.23$                 & 7.52\\ 
        SDSSJ115701     & $0.05\pm0.03$                 & 6.36\\
        NGC 3982        & $23.31\pm0.29$                & 9.01\\
        Blob            & $0.41\pm0.15$                 & 7.26\\ 
        \hline 
\end{tabular} \\
\end{table}

The velocity field in Fig.\ \ref{fig:ngc3998-hi} shows a well-defined  gradient along the major axis
of the \HI\ disk, suggesting that its kinematics are dominated by   regular rotation, at least for
radii < 10 kpc. The scale-bar (10 kpc) in the PV diagram indicates the approximate extent of the
regular rotation. Beyond this radius, the \HI\ structure becomes more filamentary. The spectra at
some locations with radii larger than 10 kpc show multiple components,  suggesting that this gas may
still be settling into a disk.  At even larger radii, some  clouds without any obvious optical
counterpart are found, the largest of which is found about 11$^\prime$ north of NGC 3982 (and named
"Blob" in Table \ref{tab:HI}). In addition, the \HI\ distribution and kinematics in the outer
regions of NGC 3982 are quite disturbed.

Faint structures and possible streams have been observed with the MegaCaM instrument on the
Canada-France-Hawaii Telescope,  as part of the CFHT Large Programme Mass
Assembly of Early-Type Galaxies with their Fine Structures (MATLAS)
\footnote{http://irfu.cea.fr/Projets/matlas/atlas3D/NGC3998.html}, which is an extension to
the study presented in \citep{Duc2015}.  However, the detection is limited by the presence of
nearby bright stars in the field. Additionally, there is no prominent optical counterpart to the
\HI\ structure.  Thus, a picture where the \HI\ of NGC~3998 was brought in by a small satellite
and is now in the process of settling in NGC~3998 appears a likely scenario.  We will come
back to this point in Sec.\ref{sec:accretion-event}. 

\begin{figure} 
    \centering 
    \includegraphics[width=\hsize]{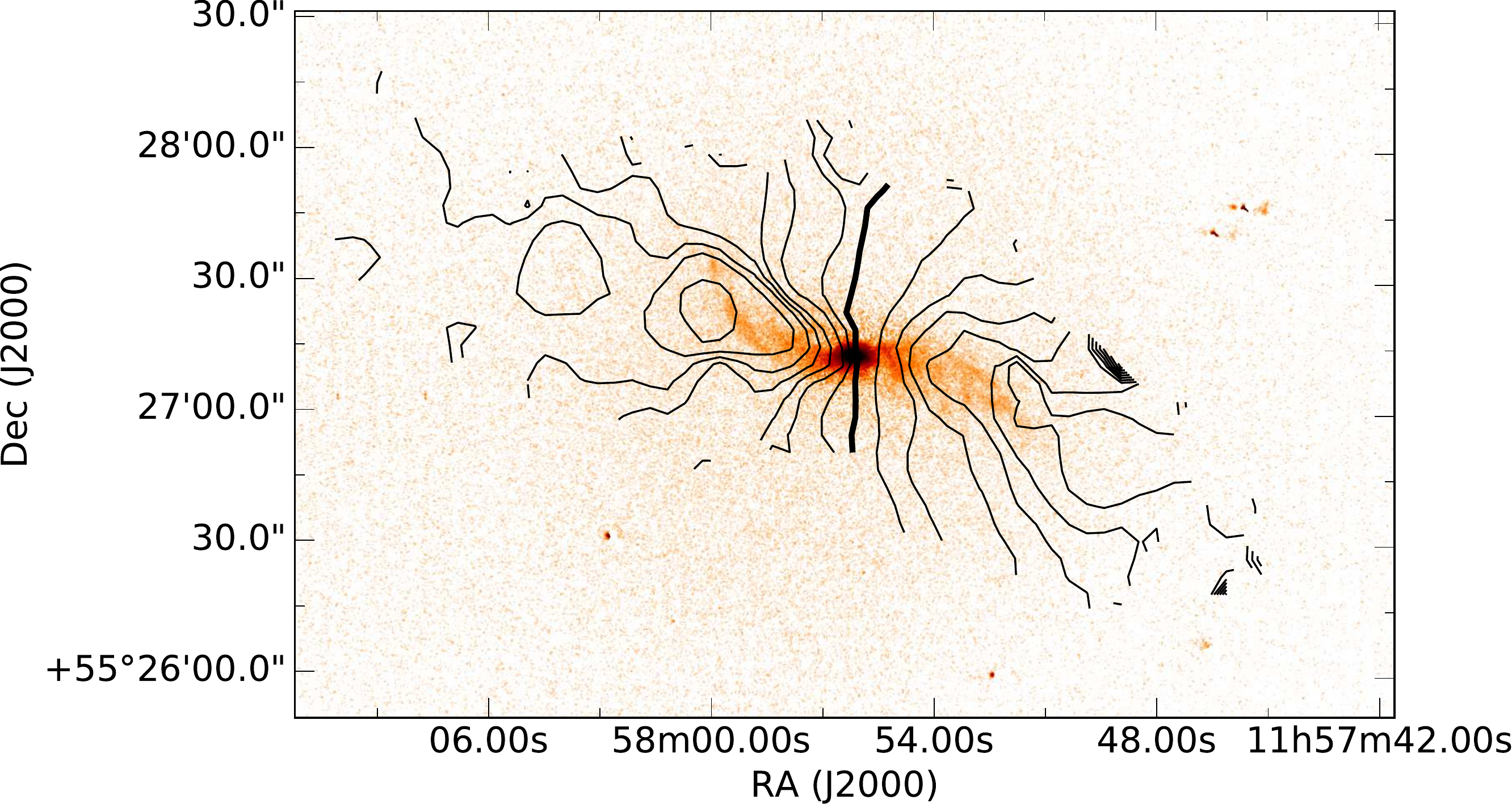}
    \caption{\label{fig:warp} Velocity contours of the inner \HI\ disk, obtained obtained from
    Gauss-Hermite fits to the 15$''$ resolution cube, superposed to the H$\alpha$ image from
    \citet{Sanchez2012}. Contour values are 800, 840, 880, \dots , 1400 \kms. The systemic velocity of
    1040 \kms\ is indicated by the thick line.} \label{velHalpha} 
\end{figure}

It is worth noting that the \HI\ and the stellar components  are arranged in a nearly polar
configuration (see  also \citealt{Knapp1985}). A perfectly polar \HI\ configuration can be
relatively stable within an oblate mass distribution (which NGC~3998 is likely to have, see
\citealt{Cappellari2013}). Using the available data on the inclination of the \HI\
\citep{denHeijer2015} and of the optical body \citep{Krajnovic2011,Cappellari2013} we find that the
position angle between the \HI\ and the symmetry plane of the stellar distribution is 80$^\circ$,
but given the uncertainties, a perfectly polar configuration cannot be entirely excluded. In Fig.\
\ref{fig:ngc3998-ha} a sketch of the orientation of the various galaxy components is presented.  The
angle (measured from North counterclockwise) between the stellar mid-plane and the \HI\ disc is
$\sim70^\circ$, (taking the PA(star) $= 135 ^\circ$ as derived from both photometric and kinematic
information,  \citealt{Krajnovic2011}, and the PA(\HI)$=65^\circ$ measured in this work). The
morphology of the inner H$\alpha$ disk strongly suggests that it has a warped structure with the
inner parts somewhat more aligned with the stellar body. To further study this, we have made a
velocity field of the \HI\ in the inner regions using the high-resolution \HI\ data cube where we
have derived the velocities using Gauss-Hermite fits in order to take into account the slight
asymmetric shape of many of the \HI\ profiles in this region (Fig.\ \ref{fig:warp}). This velocity
field shows that in the very inner region, the \HI\ disk seems to follow the warped H$\alpha$
structure.  This is visible from the curved kinematical major axis of the velocity field, and from
the fact that the kinematical minor- and major-axes are not perpendicular. This suggests  that
differential precession due to torques on the overall gas structure is occurring, in particular in
the inner regions.  We will discuss this point in Sec.\ \ref{sec:accretion-event}. 

%--------------------------------------------------------------------
\section{The nature of the radio continuum structure} 
%--------------------------------------------------------------------

The interesting new finding from our observations is the detection of  extended low
surface-brightness radio lobes representing  relatively rare structures in  low power radio sources
like NGC~3998.  The low surface brightness of the  lobes is comparable to known "remnant" lobes
(e.g. \citealt{Saripalli2012,Brienza2016}), i.e.\ structures not directly fuelled anymore by an
active nucleus. However, from a morphological point of view, the lobes of NGC~3998 have
characteristics which suggest that they are still actively fed by the nucleus. An example of this is
in the southern part which shows a collimated kpc-scale jet that may be still feeding the southern
lobe. 

The study of the  \atlas\ sample  of nearby early-type galaxies  (of which NGC~3998 is part)
suggests that this type of radio structure are genuinely rare in low power radio sources. Of the 260
nearby ($D<40$ Mpc) early-type galaxies included in the \atlas\ survey \citep[and which have 5 GHz
nuclear radio powers typically between $10^{18}$ and $10^{22}$ \WHz;][]{Nyland2016}, NGC~3998 is
among a minority (smaller than $\sim$ 10\%) of systems that are known to harbour extended radio
emission at high resolution scales \citep[see][]{Nyland2016}. Furthermore, even in the deep, lower
spatial resolution continuum images (which are more sensitive to extended emission) obtained for a
subset of these objects  from WSRT observations (Morganti et al. in prep), only  three objects out
of more than 100  observed \citep{Serra2012} were found to have extended emission on the scale of
tens of kpc, one of them being NGC~3998 (the other two NGC 3665 and NGC 5322). In this context, it
is not surprising   that NGC~3998 hosts one of the most massive black holes (above the 90-th
percentile)  among the galaxies of the \atlas\ sample and that the radio source is among the most
powerful of this sample of nearby early-type galaxies \citep{Nyland2016}. 

The case of NGC~3998 suggests that the reasons  these structures are rare are connected to their
properties. Most of the characteristics of the radio continuum structure of NGC~3998 can be
explained as being the result of poorly collimated jets.  Low power radio sources such as NGC~3998
($\log P_{\rm 1.4~GHz} = 21.63$ \WHz, i.e. well below the typical radio power of FRI radio
galaxies) are characterised by a low speed of their plasma flows. Although even in these low power
objects the jets generally seem to start with high (relativistic) velocities, they decelerate very
early in their evolution, well inside the inner  kpc \citep{Laing2014}. Thus, these jets will be
dominated by turbulence and by entrainment of  the external medium. This results in a high ratio of
non-radiating to radiating particles \citep[see
e.g.][]{Croston2008,O'Sullivan2013,Perucho2014,Laing2014}. The increased thermal component inside
the jet would further slow it down. The consequence of this would be low surface brightness
structures which would  fade away relatively quickly. In addition to this, the core prominence of
the radio continuum structure in NGC~3998 is rather extreme. 

Previous studies have shown the $\mathrm{H\alpha}$ disk to be regular
\citep{2006A&A...460..439D,Walsh2012}. Thus, there is no evidence for an interaction from the
kinematics of the ionised gas. The observed characteristics could be connected to NGC~3998 beginning
a new phase of (restarted) nuclear activity. Consequently, the observed variability of the core  may
indicate that the fuelling occurs through "discrete events", with new structures being formed and
ejected as a result. A similar hypothesis  has been recently suggested for the case of another
nearby galaxy, NGC~660 \citep{Argo2015}. Further investigating this possibility will require a
detailed study of the inner kpc involving high-resolution (VLBI) observations. 

Finally, it is important to consider the origin of the overall curved S-shape structure of the
extended lobes which (unlike the inner structure and the southern jet)  appears to be  symmetric on
the two sides.  We favour the scenario in which these curved radio lobe structures are caused by
precession, i.e.\ a change of orientation of the spin axis of the central supermassive black hole,
as is seen in many objects, e.g.\ M81 \citep{Marti-Vidal2011} and Cygnus A \citep{Steenbrugge2007}.
\citet{Steenbrugge2007} have provided a number of causes that can produce such a phenomenon. In
principle, jet axis precession due to a changing SMBH spin (due to a binary massive black hole
associated with a galaxy interaction or merger) could also cause the observed continuum
morphology. However, this would not necessarily lead to a connection between the warping disk
and the S-shaped morphology. Moreover, there is little evidence for an binary SMBH as observed
in the [OIII] emission (i.e. absence of a double peak, \citealt{Devereux2011}) and the VLBI image
shows an extension that is likely due to a jet structure \citep{Filho2002} associated with a single
SMBH. Therefore, in the case of  NGC~3998 we explore a scenario in which the cause of  the
precession is due to a misalignment between the gas channeled onto the accretion disk and the
angular momentum vector of the SMBH.  In Sec. \ref{sec:accretion-event}  we
discuss the possible origin of the warp observed in the gas, and in Sec. \ref{sec:gas-disk} we
discuss the link between this warp and the triggering and morphology of the radio continuum of
NGC~3998.

%----------------------------------------------------------------------------------------------------
\section{An accretion  event and the evolution of the gas disk \label{sec:accretion-event}}
%----------------------------------------------------------------------------------------------------

The total intensity  image and the velocity field of NGC~3998 show that the \HI\ is distributed in a
relatively regularly rotating disk  with clear signs of unsettled gas in the outer regions. Thus,
the \HI\ structure can be related to an accretion event, e.g.\ by capturing gas  from nearby
companions or by a minor merger with a gas-rich satellite. This event has brought $\sim$$10^8$
\msun\ of \HI\ into NGC~3998 without visibly affecting the optical structure of the galaxy. Small
satellite galaxies are known to be gas rich and, in fact, most of their baryonic mass is in \HI\
\citep[e.g.,][]{Maddox2015}. Accretion of a late-type satellite galaxy with a stellar mass  of
$\sim$$10^8$ \msun\ and $M_\HI/M_{\rm star} \sim 1.0$  would be sufficient to explain the observed
\HI\ mass of NGC~3998. \citet{Cappellari2013} find  the stellar mass of NGC~3998 to be
$\sim$$10^{11}$ \msun\ so the merger would have a mass ratio $\sim$1000:1 which would be consistent
with the fact that the stellar body is unperturbed. 

In NGC~3998 the inner gas disc is misaligned  with respect to the outer gas
disc and appears to be closer to the stellar mid-plane. This suggests that in the central
regions torques due to the distribution of stellar mass are important. The radius of the inner,
misaligned gas disc is $\sim$1 kpc. The rotation velocity is $\sim250\,\mathrm{km\,s^{-1}}$
\citep{denHeijer2015} and the orbital time at this radius is $\sim1.5\times10^{6}\,\mathrm{yr}$.
Therefore, we conclude that the period of time during which the gas disc has been changing
orientation within the 1 kpc radius is of the order of a few times $10^6\,\mathrm{yr}$.

\citet{Hopkins2011}   argue that stellar torques can  also produce orbit crossings and shocks in the
gas, which causes the gas to lose angular momentum and accrete onto a central black hole.  Given the
recent revival of the nuclear activity in NGC~3998 this poses the interesting question whether the
start of the alignment of the gas disk is connected to the current episode of nuclear activity in
NGC~3998. 

The warped inner disk fits in a scenario of a minor gaseous merger. \citet{vandeVoort2015} have
recently presented numerical simulations of gas accretion by a large early-type galaxy  in a
cosmological setting. The simulations suggest  that it is possible that perfect alignment between a
large-scale  gas disk and the stellar body can take a long time to achieve, i.e.\  $\gtrsim$ 1
Gyr.  This is because as long as gas with misaligned angular momentum is being accreted, the overall
gas structure will remain misaligned. In  models such as those presented by \citet{vandeVoort2015},
the realignment starts in the inner regions, where the dynamical timescales are  shorter. This
implies that when realignment sets in, the inner gas disk should become warped first.   A situation
as described by \citet{vandeVoort2015} may be occurring in NGC~3998.  The gas
structure beyond a disk radius of approximately $\sim 10\,\mathrm{kpc}$is still fairly
irregular; this implies that accretion of gas is still occurring. On the other hand, given the
extent of that part of the \HI\ structure which appears in fairly regular rotation, the
accretion in NGC~3998 must have started about a Gyr ago, assuming it takes several orbital
periods for a regular gas disk to form.  Due to the effects discussed by \citet{vandeVoort2015},
the inner disk has started to align only recently.

%----------------------------------------------------------------------------------------------------
\section{Gas disk and the radio continuum emission \label{sec:gas-disk}}
%----------------------------------------------------------------------------------------------------

In a number of radio galaxies, the presence of large-scale  \HI\ structures have allowed us to
compare the  timescale of the gas accretion/merger with the age of the radio sources  (see e.g.\
B2~0648+27, \citealt{Emonts2006};  Cen~A, B2~0258+35, \citealt{Struve2010a,Struve2010b}; and
PKS~1718-63, \citealt{Maccagni2014}). In all these cases,   a typical age of the radio source of
$10^6$ -- $10^7$ yr is found and a timescale for the accretion/merging of $10^8$ - $10^9$ yr. This
implies an  extremely large delay between the start of the accretion and the onset of nuclear
activity and this makes  it difficult to identify a direct link between the two phenomena.

For NGC~3998 the multi-frequency radio data necessary to derive the age of the radio source are note
available.  Thus, we have estimated the age of the radio continuum emission by making assumptions
about the jet speed and the time needed to build the observed lobes.  Even for low-power jets, it
has been claimed that their initial velocity is relativistic, or close to. However, these jets are
observed to  rapidly decelerate  to sub-relativistic speeds, and this is likely happening well
inside the inner few kiloparsecs. The best example is M84 ($P_{\rm 1.4 GHz} \sim 10^{23.4}$ \WHz)
for which \citet{Laing2014} suggest that the jets are initially relativistic, but the deceleration
scale is estimated to be only 1.8 kpc. Beyond  that radius the velocities must be quite small (i.e.\
below $c/10$).  The lowest-power radio galaxy where the characteristic structure of deceleration
from relativistic speeds is observed is B2~1122+39 \citep{Laing1999}.  

Direct determination of jet speeds has been done only in a few cases \citep[e.g.\  NGC3801 and Cen
A;][]{Croston2007,Croston2009}. In those cases, the values (ranging between 850 and 2600 \kms\
respectively) have been derived from the presence of X-ray shells around the radio emission and
deriving the Mach number of the radio flow. In NGC~3998, the radio emission is more relaxed so we do
not consider the flow to have a supersonic speed (i.e.\ no shock is observed).  Sound speed of the
jets have been given in \citet{Morganti1987} for a group of low-power radio galaxies and these range
between  700 and 1700 \kms.  Thus, for the large-scale continuum structures of NGC~3998 we assume a
low jet speed between  1 and 2 $\times 10^3$ \kms. We also assume a distance from the centre of the
galaxy to the projected edge of the radio continuum emission of $10$ kpc. The derived timescales
$\tau_\mathrm{jet}$ are between $5 \times 10^6 $ and $10^7$ yr and similar to those derived for,
e.g.,  NGC~3801 and the inner lobes of Cen A \citep{Croston2007,Croston2009}.

As in the other objects mentioned above, the time since the accretion event  started is much longer
than the age of the radio source. However, we note that our estimate of the age of the radio
emission is similar to the timescale we estimated for the inner warp in NGC~3998 to develop.
Moreover, we also note  the morphological similarity between the warp of the inner gas  disk and
the S-shaped radio continuum emission  (see Fig.\ \ref{fig:warp}).  Based on this, we suggest that
the start of the realignment of the inner gas disk could be indeed an indication of the moment when
the gas has reached  the central regions. Additionally, the effects of the  stellar torques  provide
the necessary loss of angular momentum which would  cause   gas clouds  to fall toward the SMBH,
fuel it and make it active. In this scenario, the connection between the
warping of the gas structure and the morphology of the radio lobes would be that the SMBH spin
and, therefore, the jet axis are adapting to the changing angular momentum of the accreting
gas. Indeed, such jet precession is expected to generate the radio continuum morphology
observed in systems such as Mrk 6 \citep{2006ApJ...652..177K} and NGC~326 \citep{1978Natur.276..588E}.

%----------------------------------------------------------------------------------------------------
\section{Energetics and triggering}
%----------------------------------------------------------------------------------------------------

Based on its optical emission line ratios, NGC~3998 is classified as a LINER
\citep{Ho1997}. As is typical for this type of low-luminosity AGN, the accretion rate of the
SMBH in the nucleus of NGC~3998 is believed to be quite low. \citet{Eracleous2010} reported an
Eddington ratio of $4 \times 10^{-4}$, indicating inefficient SMBH accretion well below the
Eddington luminosity, consistent with the classification of NGC~3998 as a Low Excitation Radio
Galaxy \citep[LERG;][]{Best2012}. The inefficient accretion characteristic of LERGs may be driven by
gas cooling from the hot galactic halo supplying fuel to the SMBH  via Bondi accretion
\citep{Croton2006,Allen2006,Hardcastle2007,Balmaverde2008}, or perhaps by chaotic accretion of
turbulent clouds \citep[e.g.,][and references therein]{Gaspari2015}.

Deeper insights into the underlying accretion physics in the nucleus of NGC~3998 may be gained by
examining the efficiency of the radio lobes relative to the mass accretion rate. Previous studies
have reported jet efficiencies of a few percent based on studies of the correlation between radio
jet power and the Bondi mass accretion rate ($\dot{m}_{\mathrm{B}}$, \citealt{Bondi1952}) for
samples of nearby, inefficiently-accreting radio AGNs
\citep{Allen2006,Balmaverde2008,Russell2013,Nemmen2015}.

The radio jet power, $P_{\mathrm{jet}}$, may be estimated based on the results of scaling relations
between 1.4~GHz radio luminosity and the jet power provided. Using the relation given in
\citet{Cavagnolo2010} and the total 1.4~GHz luminosity of NGC~3998 of
$6.15\times10^{37}\,\mathrm{erg~s^{-1}}$, $P_{\mathrm{jet}} \sim 1.64 \times 10^{42}\,
\mathrm{erg~s^{-1}}$.  Based on the relation between $P_{\mathrm{jet}}$ and the ``Bondi power'',
$P_{\mathrm{B}}=\dot{m}_{\mathrm{B}} c^2$ provided in \citet{Balmaverde2008} and our estimate of the
jet power in NGC~3998, we estimate a Bondi accretion rate in NGC~3998 of $2\times
10^{-3}\,\mathrm{M_\odot \, yr^{-1}}$. This accretion rate\footnote{We emphasize that this is only a
rough estimate since the intrinsic scatter in both the $P_{\mathrm{jet}}-P_{\mathrm{radio}}$ and
$P_{\mathrm{jet}}-P_{\mathrm{acc}}$ relation are both $\sim 0.7$~dex
\citep{Balmaverde2008,Cavagnolo2010,Nemmen2015}.} is consistent with that estimated in
\citet{Devereux2011} based on the observed $\mathrm{H\alpha}$ and X-ray emission in the NGC~3998
nucleus of $\dot{m} \sim 3.6 \times 10^{-3} \,\mathrm{M_\odot\,yr^{-1}}$. Thus, about 1\% of the
mass inflow rate is converted in jet mechanical energy in NGC~3998, consistent with other low-power
radio AGNs \citep{Balmaverde2008,Nemmen2015}.

The presence of clouds either from the cooling of a hot halo or from the accretion of gas is likely
quite common in early-type galaxies \citep{Negri2015}. These galaxies are now recognised to often
have a rich ISM and a relatively large reservoir of gas
\citep{Mathews2003,Morganti2006,Young2011,Serra2012}.  In fact, the presence of clouds of cold gas
in unsettled/chaotic orbits, possibly triggering/fuelling the SMBH and maintaining a galaxy's
nuclear activity, has been suggested for other (radio) AGN from, e.g.,  \HI\ absorption features
\citep[e.g.][and refs.  therein]{Maccagni2014}  and by the high velocity dispersion of the molecular
gas in the central regions of some low-luminosity AGN
\citep{MuellerSanchez2006,Neumayer2007,MuellerSanchez2013,Mezcua2015,Maccagni2016}. 

As mentioned above, it is possible that the torques responsible for realigning the gas disk  in the
central regions, by which some gas clouds lose their angular momentum and fall into the SMBH, is the
mechanism actually responsible for providing discreet "feeding events" for the AGN. This  can also
explain the variability observed in radio (see Sec.\ 4.1).  Furthermore, we suggest that low-power
radio continuum lobes, like the ones we see in NGC~3998, are short-lived structures. This is due to
the low speed of the jets and the consequent strong turbulence and entrainment of material from the
interstellar/intergalactic medium (ISM/IGM).  Thus,  with NGC~3998 we have caught a galaxy not too
late after the event that  fuelling the SMBH has occurred while the radio lobes were still visible. 

%--------------------------------------------------------------------
\section{Summary and Conclusions}

We have presented deep radio continuum and \HI\   observations   of the galaxy NGC~3998, performed
with the WSRT. The good sensitivity of our observations has revealed two important aspects of
NGC~3998 and the surrounding environment. Our conclusions are as follows:

\begin{enumerate}
\item{The radio continuum observations reveal two low-surface brightness, S-shaped lobes with a
total extent  of about 20 kpc. In galaxies of similar low radio power as NGC~3998 such structures
are only seldom seen. The radio continuum morphology is extremely core dominated ($S_{\rm
core}/S_{\rm ext} = 8$) and asymmetric on small scales with a 1.5 kpc jet only on one side.  We
suggest the radio emission to be the result of poorly collimated jets characterised by a low
velocity of the plasma flow. These structures are known to be dominated by turbulence and
entrainment of material from the ISM/IGM, which would further slow down the plasma flow. Hence, they
will fade away on relatively short timescales, therefore explaining the rarity of low surface
brightness radio lobes in galaxies like NGC~3998. }

\item{The \HI\ is distributed in a disk which is almost, but possibly
not quite, polar with respect to the stellar distribution. This disk is regularly rotating out to
$\sim$10 kpc from the centre, but characterised by more irregular kinematics in the outer parts.
The \HI\ in NGC~3998 is likely the result of the merger with a  relatively small galaxy ($M_{\rm
star} \sim 10^8$ \msun) which occurred about 1 Gyr ago and which has left no signature in the
optical appearance of the galaxy. }

\item{ The inner \HI\ disk, together with the inner disk of ionised gas, is warping in a way mirroring the
S-shape of the radio lobes and we suggest a connection between the two.  We argue that the gas disk
in the central regions has only recently (a few $\times 10^6$ yr ago)  started to align with the
stellar major axis. The effect of the stellar torques that case the warping of the disk could also
cause the gas clouds in the central regions to lose angular momentum and fall into the super-massive
black hole.  We suggest that  this mechanism may have triggered the radio continuum source,
something we infer has happened on comparably short timescales. In this scenario, the S-shape  of
the radio structure would be due to the jet axis adapting to the changing angular momentum axis of
the warping gas disk.}
\end{enumerate}

We predict additional large-scale, diffuse radio structures similar to those in NGC~3998 will be
detected in other galaxies by new, deeper observations that are sensitive to low-surface-brightness
emission. Future studies are planned to further explore the properties of the newly-discovered
diffuse radio lobes of NGC~3998.  These include observations with the Low Frequency Array
\citep{vanHaarlem2013}, the Giant Metrewave Radio Telescope, and the Karl G. Jansky Very Large
Array that will allow a detailed analysis of the radio spectral indices of the continuum
emission of the diffuse lobes of NGC~3998.

%-----------------------------------------------------------------------------------------------
\begin{acknowledgements}

We thank Freeke van der Voort for useful discussions and suggestions during the preparation of this
paper. We also thank J. R. S\'anchez-Gallego for providing the H$\alpha$ image used in Fig.\
\ref{velHalpha}.  The research leading to these results has received funding from the European
Research Council under the European Union's Seventh Framework Programme (FP/2007-2013) / ERC
Advanced Grant RADIOLIFE-320745. The Westerbork Synthesis Radio telescope is operated by the
Netherlands Institute for Radio Astronomy (ASTRON) with support of the Netherlands Foundation for
Scientific Research (NWO). We thank the referee for the constructive comments which have helped
improve this paper.  

\end{acknowledgements}

\end{document}